\documentclass[pra,aps,twocolumn,showpacs,10pt,floatfix]{revtex4-1}
\usepackage{amsmath,amssymb,graphicx,bm,color,ulem}
\usepackage{amsfonts,amsthm}

\begin{document}

\title{Clarification of the transverse orbital angular momentum of \\ spatiotemporal optical vortices}

\author{Miguel A. Porras}
\email{miguelangel.porras@upm.es}
\affiliation{Grupo de Sistemas Complejos, ETSIME, Universidad Politécnica de Madrid, Rios Rosas 21, 28003 Madrid, Spain}

\begin{abstract}
Advances in the generation and the application of spatiotemporal optical vortices (STOV) are proceeding fast, but fundamental aspects of their nature remain obscure. Phys. Rev. A 107, L031501 (2023) (PRA) and Prog. Electromagn. Res. 177, 95 (2023) (PIER) provide contradictory results on the transverse orbital angular momentum (OAM) carried by STOVs. We show that the results by Porras in PIER and by Bliokh in PRA refer to different STOVs and are all correct. In PIER, STOVs are elliptical at given cross section and time, or in space-time, but not in three-dimensional space. In PRA, STOVs are elliptical in space but not in space-time. This is evidenced from two dual, equivalent theories on the transverse OAM where a wave packet is seen in space-time evolving with propagation distance or in space evolving in time, that accounts for all values of the total, intrinsic and extrinsic OAM in PIERS and PRA. However, the intrinsic OAM with respect to the photon wave function center in PRA is not generally conserved, which advocates for the energy center in PIER as the STOV center. We argue that STOVs are generated in experiments to purportedly have elliptical symmetry in space-time. The values provided in PIER should then be taken as the reference for elliptical STOVs, and the theory therein to evaluate the transverse OAM of other wave packets. Hancock {\it et al.} in Phys. Rev. Lett. 127, 193901 (2021) and Phys. Rev. X. 14, 011031 (2024) erroneously attribute the transverse OAM of elliptical STOVs in space to the elliptical STOVs in space-time they consider theoretically and can generate in their experiments.
\end{abstract}

\maketitle

\section{Introduction}

Spatiotemporal optical vortices (STOVs), spatiotemporally structured fields with a line phase singularity transverse to their propagation direction, are rapidly gaining attention among the wide variety of structured light forms. Starting with their controlled generation with $4f$ pulse shapers, either with spiral phase plates \cite{HANCOCK_OPTICA_2019} or spatial light modulators \cite{CHONG_NATPHOT_2020}, new methods have been developed based on nanostructures \cite{HUANG_PRB_2023}, photonic crystals \cite{WANG_OPTICA_2021}, and more recently on the use of a single lens adequately illuminated \cite{PORRAS_PRA_2024}. These new wave objects are finding application in particle manipulation \cite{STILGOE_NATPHOT_2022}, as information carriers \cite{WAN_ELIGHT_2022,HUANG_SCI_2024}, and are used as driving fields for the generation of second harmonic \cite{GUI_NATPHOT_2022} and high-order harmonic STOVs \cite{FANG_PRL_2021}. For a review see, for instance, Ref. \cite{WAN_ELIGHT_2023}.

However, fundamental properties of STOVs such as the nature and amount of their transverse orbital angular momentum (OAM) are still problematic  because different authors present different results \cite{PORRAS_PIERS,BLIOKH_PRA_2023,BLIOKH_PRL_2021,BLIOKH_PRA_2012}. In this paper we show that the different results actually refer to different STOVs, and therefore do not contradict each other. In Ref. \cite{PORRAS_PIERS}, STOVs with elliptical symmetry at a transversal plane and in time, or in space-time, are considered, which are not elliptical when observed in three-dimensional space. Ref. \cite{BLIOKH_PRA_2023,BLIOKH_PRL_2021,BLIOKH_PRA_2012} considers STOVs with elliptical symmetry in three-dimensional space, which have not elliptical symmetry in space-time. ``Elliptical symmetry" refers indistinctly to the intensity (flux of the Poyinting vector) or the energy density as both are proportional under the paraxial conditions considered here. We analyze both families and conclude that the different values reported in \cite{PORRAS_PIERS} and \cite{BLIOKH_PRA_2023,BLIOKH_PRL_2021,BLIOKH_PRA_2012} of the transverse OAM, and its extrinsic and extrinsic parts, are all correct.

The elliptically symmetric STOVs in space-time are more easily analyzed from a theory of transverse OAM of wave packets in which they are specified in space-time at a certain transversal plane and evolve from plane to plane, as in \cite{PORRAS_PIERS}, which we will refer to as ``space-time formulation." The OAM of a general wave packet is evaluated by as the time-integrated angular momentum flux through a transversal plane. The elliptically symmetric STOVs in space are more naturally analyzed from a theory where the fields are specified in space and evolve in time. We develop here such a theory, referred to here as ``spatial formulation", where the OAM of a general wave packet is evaluated as the angular momentum density integrated in space. The equivalence of these two formulations is established theoretically and checked numerically.

For STOVs propagating along the $z$ direction and phase line singularity along the $y$ direction, we evaluate the transverse OAM about a fixed $y$ axis crossing at an instant of time their center of symmetry, and extract its extrinsic and intrinsic parts taking the center of the energy as the center of the STOV. For elliptical STOVs in space-time we recover the results in Ref. \cite{PORRAS_PIERS} using both the space-time and spatial  formulations that the transverse OAM vanishes because the extrinsic and intrinsic OAM are opposite, or expressed per unit energy $W$,
\begin{equation}\label{UNO}
\frac{J_y}{W}=0, \frac{J_y^{(e)}}{W}=-\frac{l}{2\omega_0}\gamma, \frac{J_y^{(i)}}{W}=\frac{l}{2\omega_0}\gamma ,
\end{equation}
for a STOV carrier frequency $\omega_0$, topological charge $l$, and ellipticity $\gamma$. In passing the spatial formulation allows us to visualize the non-ellipticity in three-dimensional space of elliptical STOVs in space-time, which intuitively explains the extrinsic OAM. For elliptical STOVs in space we recover the results in Ref. \cite{BLIOKH_PRA_2023}, also from both formulations, that the transverse OAM is equal to the intrinsic transverse OAM because the extrinsic part vanishes (when the center of the energy is taken as the STOV center), i.e.,
\begin{equation} \label{DOS}
\frac{J_y}{W}=\frac{l}{2\omega_0}\gamma, \frac{J_y^{(e)}}{W}=0, \frac{J_y^{(i)}}{W}=\frac{l}{2\omega_0}\gamma .
\end{equation}
In passing, the space-time formulation allows us to visualize the non-ellipticity of these STOVs in space-time, confirming that the two families of STOVs in \cite{PORRAS_PIERS} and \cite{BLIOKH_PRA_2023,BLIOKH_PRL_2021,BLIOKH_PRA_2012} are different. The spatial formulation trivially allows us to shift from the center of the energy to the photon wave function center as alternative definition of STOV center to recover the results in \cite{BLIOKH_PRA_2023,BLIOKH_PRL_2021,BLIOKH_PRA_2012} that
\begin{equation}\label{TRES}
\frac{J_y}{W}=\frac{l}{2\omega_0}\gamma, \frac{J_y^{(e)\prime}}{W}=-\frac{l}{2\omega_0}\frac{1}{\gamma}, \frac{J_y^{(i)\prime}}{W}=\frac{l}{2\omega_0}\left(\gamma+ \frac{1}{\gamma}\right) .
\end{equation}
Nevertheless, the extrinsic and intrinsic transverse OAM with respect to the photon wave function center are not conserved in general, as seen here with an example. The non-conservation is not a failure {\it per se} as there no exist general laws of conservation of the intrinsic and extrinsic OAM, particularly for STOVs whose elliptical symmetry in space-time or in space is lost on propagation. The conservation with respect to the energy center \cite{PORRAS_PIERS} makes the choice of the energy center clearly preferable.

On the other hand, Hancock {\it et al.} in Ref. \cite{HANCOCK_PRL} first considered that STOVs have purely intrinsic transverse OAM, as longitudinal vortices, on the basis of the expectation values of certain operators for the extrinsic and intrinsic parts. It has been demonstrated that these operators are not Hermitian and therefore cannot represent any physical magnitude \cite{PORRAS_COMMENT_PRL}. The existence of extrinsic transverse OAM is later recognized in \cite{MILCHBERG_PRX}. However, the authors believe that the elliptical STOVs in space-time \cite{HANCOCK_PRL} (those they can generate in their experiments), are also elliptical in space, attributing erroneously to the former the OAM properties of the latter \cite{MILCHBERG_PRX}.

The space-time formulation is indeed closer to experiments with ultrashort pulses, and STOVs with elliptical symmetry in space-time are closer to the STOVs generated in experiments. Pulse and beam shaping techniques are usually aimed at producing a wave packet with desired characteristics at a transversal section and in time, as elliptical, or purportedly elliptical STOVs at a focal plane. Also, characterization techniques allows to retrieve optical fields with spatial resolution at a transversal section and in time. It is no coincidence that these optical fields are referred to as spatiotemporal fields, and that theories have been developed on their spatiotemporal couplings.
This is why we believe Eqs. (\ref{UNO}) should be taken as a reference for canonical STOVs, and the space-time formulation when evaluating the transverse OAM of more general spatiotemporal fields.

Of course, the transverse OAM and its extrinsic part depend on the choice of the transverse axis, but in the end all this debate about the OAM about the STOV center of symmetry has led to firmly establish the values of the total and intrinsic part for the various types of canonical STOVs, which will have implications on how STOVs interact with particles \cite{STILGOE_NATPHOT_2022}, atoms and light \cite{GUI_NATPHOT_2022,FANG_PRL_2021}, and on the way, to unveil details of their innermost structure that have gone unnoticed.

\section{Preliminaries}\label{PRELIMINAIRES}

Let us begin with the physical magnitudes that are conserved according to Maxwell equations in free space. Their conservation is expressed mathematically by means of continuity equations. Conservation of energy is expressed by the continuity equation $\partial_t w + \partial_i S_i=0$ (repeated subindexes are summed over all its values $i=x,y,z$), where $w=(\varepsilon_0 E_iE_i + \mu_0^{-1}B_iB_i)/2$ is the energy density, and $S_i = \mu_0^{-1}\epsilon_{ijk} E_jB_k$ is the Poynting vector or energy flux density ($\epsilon_{ijk}$ is the permutation symbol of values $\epsilon_{ijk}=+1$ if $ijk = 123, 312, 231$,  $\epsilon_{ijk}=-1$ if $ijk = 321, 132, 213$, and zero otherwise). Conservation of momentum is expressed by $\partial_t p_i +\partial_j T_{ij} =0$, where $p_i=S_i/c^2$ is the momentum density and $T_{ij}=(1/2) \delta_{ij}(\varepsilon_0 E_kE_k + \mu_0^{-1}B_kB_k) - \varepsilon_0 E_i E_j - \mu_0^{-1}B_iB_j $ is the momentum flux density ($\delta_{ij}$ is the Kronecker delta). $T_{ij}$ is the flux of the $i$ component of the momentum per unit surface perpendicular to the $j$ direction.

The conservation of angular momentum implies the third continuity equation $\partial_t j_i + \partial_m M_{im}$, where $j_i =\epsilon_{ikl}x_kp_l$ is the angular momentum density and $M_{im}=\epsilon_{ikl}x_kT_{lm}$ is the angular momentum flux density. $M_{im}$ is the flux of the $i$ component of angular momentum per unit surface perpendicular to the $m$ direction, with units of angular momentum per unit time and unit area.

Any of these continuity equations can be expressed in integral form. We will need the integral forms for the energy and angular momentum. Integration of the respective continuity equations over an arbitrary volume and use of the divergence theorem leads to
\begin{equation}\label{CONTINUITY}
\frac{d}{dt}\int_V w dV = -\oint_S S_m ds_m,\quad \frac{d}{dt}\int_V j_i = - \oint_S M_{im} ds_m,
\end{equation}
expressing that the variation of the energy and angular momentum in a volume $V$ is only due to the inward energy and momentum flux across its boundary $S$.

\begin{figure}[h!]
\begin{center}
\includegraphics*[height=3cm]{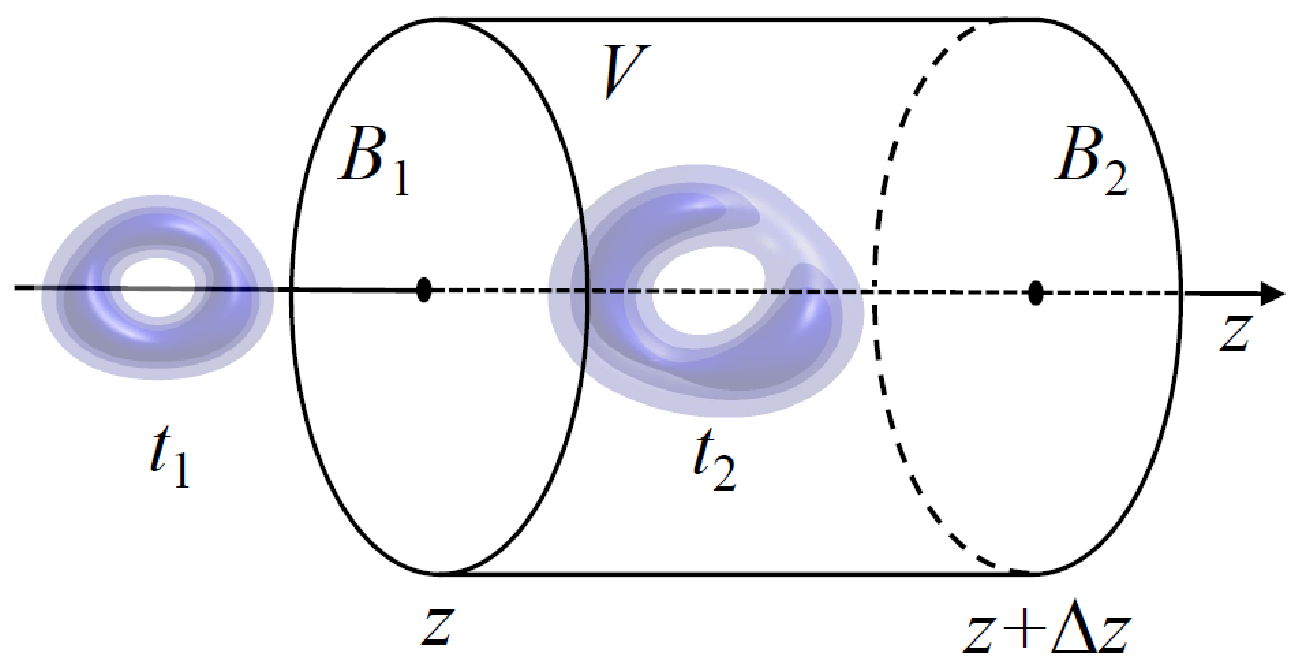}
\end{center}
\caption{\label{Fig1} Cylindrical volume for the application of the divergence theorem to the energy and angular momentum densities to a wave packet localized in three dimensions.}
\end{figure}

Consider now the particular case of a wave packet localized in the three spatial dimensions and propagating at the velocity of light towards positive $z$. We take a sufficiently long and wide cylinder $V$ as in Fig. \ref{Fig1} between the planes $z$ and $z+\Delta z$ such that the whole wave packet is outside $V$ at a time $t_1$ and the whole wave packet is inside $V$ at $t_2$. The continuity equations in Eqs. (\ref{CONTINUITY}) for this surface yields, upon integration in time between $t_1$ and $t_2$,
\begin{equation}
%\begin{split}
\left.\int_V \!\!\!w dV \right]^{t_2}_{t_1} \! = \!\int_{t_1}^{t_2}\!\!\!\int_{B_1}\!\!\! S_z ds_z dt ,\,\,
\left.\int_V \!\!\!j_i dV \right]^{t_2}_{t_1}\!  = \!\int_{t_1}^{t_2}\!\!\!\int_{B_1}\!\!\! M_{iz} ds_z dt, \nonumber
%\end{split}
\end{equation}
since there has been flux only across the basis $B_1$. The minus signs in Eqs. (\ref{CONTINUITY}) cancel when taking $ds_z$ towards positive $z$. Since the whole wave packet is outside $V$ at $t_1$, the volume integrals evaluated at $t_1$ vanish. Since the whole wave packet is inside $V$ at $t_2$, the volume $V$ of the integrals at $t_2$ can be replaced by $\mathbb{R}^3$ and identified with the energy $W$ and angular momentum $J_i$ of the wave packet. For sufficiently large radius, the integration over $B_1$ can be extended to the whole transversal plane. Also, the temporal limits $t_1$ and $t_2$ in the right hand side can be extended to all times since there is only flux in the time interval between $t_1$ and $t_2$. We then obtain two alternative expressions for the energy and the angular momentum carried by the wave packet:
\begin{equation}
W = \!\!\int\!\! w dV = \!\!\int\!\! S_z ds_z dt, \quad \! J_i = \!\! \int \!\!\! j_i dV = \!\!\int\!\! M_{iz} d s_z dt ,
\end{equation}
where the infinite limits in all integrals are omitted for brevity. The volume integrals for the energy and angular momentum are conserved in time $t$. The corresponding flux integrals are conserved on propagation along $z$.

To date, STOVs created in laboratories propagate paraxially and are quasi-monochromatic or narrow-band, the precise meaning of which is that they contain many optical cycles, with typical of durations from several tens to hundreds of femtoseconds. Under these conditions electromagnetic fields can be constructed from a paraxial and quasi-monochromatic scalar field $\psi e^{i(k_0z-\omega_0t)}$ of carrier frequency $\omega_0$ and wave number $k_0=\omega_0/c$ as \cite{LAX}
\begin{eqnarray}\label{FIELDSX}
E_x &=& {\rm Re}\left\{\psi e^{i(k_0z- \omega_0t)}\right\},  B_y={\rm Re}\left\{\frac{1}{c}\psi e^{i(k_0z- \omega_0t)}\right\}, \nonumber\\
& & \quad \quad E_z= {\rm Re}\left\{\frac{i}{k_0}\partial_x \psi e^{i(k_0z- \omega_0t)}\right\}, \\
& & \quad \quad B_z = {\rm Re}\left\{\frac{i}{k_0c}\partial_y \psi  e^{i(k_0z- \omega_0t)}\right\}. \nonumber
\end{eqnarray}
for polarization along $x$, and
\begin{eqnarray}\label{FIELDSY}
E_y &=& {\rm Re}\left\{\psi e^{i(k_0z- \omega_0t)}\right\}, B_x=-{\rm Re}\left\{\frac{1}{c}\psi e^{i(k_0z- \omega_0t)}\right\}, \nonumber \\
& & \quad \quad E_z ={\rm Re}\left\{\frac{i}{k_0}\partial_y \psi e^{i(k_0z- \omega_0t)}\right\}, \\
& & \quad \,\,\,\, B_z=-{\rm Re}\left\{\frac{i}{k_0c}\partial_x \psi  e^{i(k_0z- \omega_0t)}\right\}. \nonumber
\end{eqnarray}
for polarization along $y$.

In Ref. \cite{PORRAS_PIERS}, the fields are specified and analyzed in space-time at a given transversal plane $z$ as $\psi(x,y,t',z)$, where $t'=t-z/c$ is the local time at the plane $z$, and where the argument $z$ takes into account the changes of the field on propagation distance. This is also the point of view in most of experimental and theoretical works on ultrashort pulses. In Ref. \cite{BLIOKH_PRA_2023}, the fields are instead specified and analyzed in three-dimensional space at a given time as $\psi(x,y,z',t)$, where $z'=z-ct$ is used instead of $z$ to eliminate the fast variation in time of a wave packet traveling at the velocity of light $c$, and the argument $t$ takes into account the remaining changes of the field from time to time.

In the following we recall and improve the space-time formulation for the determination of the transverse OAM in Ref. \cite{PORRAS_PIERS} based on a wave packet specified in space-time $(x,y,t')$ at a plane $z$, and develop a new spatial formulation based on a wave packet specified in $(x,y,z')$ at an instant of time $t$. Note that the equations that determine axial dynamics in $z$ of $\psi(x,y,t',z)$ and the temporal dynamics in $t$ of $\psi(x,y,z',t)$ are not needed at the moment as it suffices to evaluate the transverse OAM at a single plane in the space-time formulation due to its conservation in $z$, and it suffices to evaluate the transverse OAM at a single instant of time in the spatial formulation due to its conservation in $t$. Those dynamical equations will only be needed when comparing the results from the two formulations.

\section{Space-time formulation}

The energy and the angular momentum about a transverse axis, say $y$, through the origin O of a wave packet propagating along the $z$ direction at speed $c$ are conveniently evaluated here as
\begin{equation}
W=\int \langle S_z\rangle d\vec x_\perp dt',\quad J_y = \int \langle M_{yz} \rangle d\vec x_\perp dt',
\end{equation}
where $ds_z=dxdy\equiv d\vec x_\perp$, and we have averaged over the fast oscillations of the quasi-monochromatic field. We have taken into account that integration to all $t$ at given $z$ is the same as integration to all $t'$. The energy flux density is $S_z=\mu_0^{-1}(E_xB_y - E_y B_x)$, and $M_{yz} = zT_{xz}- xT_{zz}$, where $T_{xz}=-\varepsilon_0 E_xE_z - \mu_0^{-1} B_xB_z$ and $T_{zz}=[\varepsilon_0(E_x^2+E_y^2-E_z^2)+ \mu_0^{-1}(B_x^2+B_y^2-B_z^2)]/2$ are the $x$ and $z$ components of the momentum flux density along $z$.

The averaged $S_z$ with the fields in either Eqs. (\ref{FIELDSX}) or Eqs. (\ref{FIELDSY}) yields $\langle S_z\rangle =\varepsilon_0c|\psi|^2/2$ and the energy as $W=\int \langle S_z\rangle d \vec x_\perp dt'$, or
\begin{equation}\label{EO}
W=\frac{\varepsilon_0 c}{2} \int |\psi|^2 d\vec x_\perp dt'.
\end{equation}
Similarly, the cycle-averaged angular momentum flux density is
\begin{equation}\label{MYZ}
\langle M_{yz} \rangle = z\langle T_{xz}\rangle -x\langle T_{zz}\rangle,
\end{equation}
where
\begin{equation}\label{TXZ}
  \langle T_{xz}\rangle = \frac{\varepsilon_0}{2k_0} {\rm Im}\left\{ \psi^\star \partial_x\psi\right\}, \quad \langle T_{zz}\rangle=\frac{\varepsilon_0}{2}|\psi|^2\,.
\end{equation}
The transverse angular momentum about the origin O is then
\begin{equation}\label{JYO}
J_y=\frac{\varepsilon_0 z}{2k_0}\int {\rm Im}\left\{\psi^\star \partial_x \psi\right\}  d\vec{x}_\perp dt' -\frac{\varepsilon_0}{2}\int |\psi|^2 x d\vec{x}_\perp dt'.
\end{equation}
We note that the angular momentum flux density in Eq. (\ref{MYZ}) may contain a small contribution from spin angular momentum \cite{BARNETT_JOPT_2002} due to the slight elliptical polarization. For paraxial fields, the axial components and the contribution of the slight elliptical polarization to spin angular momentum are very small, and in any case this contribution integrates to zero in Eq. (\ref{JYO}) \cite{BLIOKH_PRL_2021}, which we will call transverse OAM.

We can extract the extrinsic part of the transverse OAM as follows. At the plane $z$ the center of the energy flux is transversally located at
\begin{equation}\label{CENTERXO}
\verb"x"_{\rm CE}(z) = \frac{1}{W}\int \langle S_z\rangle x d\vec{x}_\perp dt'=\frac{\int |\psi|^2 x d\vec{x}_\perp dt'}{\int |\psi|^2 d\vec{x}_\perp dt'},
\end{equation}
a similar equation for $\verb"y"_{\rm CE}(z)$, and at $z+c[t'-\verb"t"'_{\rm CE}(z)]$ as the local time $t'$ runs at that plane. The time
\begin{equation}\label{CENTERTO}
\verb"t"'_{\rm CE}(z) = \frac{1}{W}\int \langle S_z\rangle t' d\vec{x}_\perp dt'= \frac{\int |\psi|^2 t' d\vec{x}_\perp dt'}{\int |\psi|^2 d\vec{x}_\perp dt'}
\end{equation}
accounts for the possibility that the temporal center may not be $t'=0$ for a general wave packet without particular symmetries. We stress that these parameters are not related to the spatial structure of the wave packet in $(x,y,z)$ but to what happens at a transversal plane in $(x,y,t')$. We then write the extrinsic transverse OAM as
\begin{eqnarray}
  J_y^{(e)} &=& \int \{z+c[t'-\verb"t"'_{\rm CE}(z)]\}\langle T_{xz}\rangle  d\vec x_\perp dt' \nonumber \\
  &-&\verb"x"_{\rm CE}(z)\int \langle T_{zz}\rangle d\vec x_\perp dt',
\end{eqnarray}
and using Eqs. (\ref{TXZ}) and (\ref{CENTERXO}), as
\begin{eqnarray}\label{JYEO}
J_y^{(e)}&=&\frac{\varepsilon_0 z}{2k_0}\int {\rm Im}\left\{\psi^\star \partial_x \psi\right\} d\vec{x}_\perp dt'- \frac{\varepsilon_0}{2} \int |\psi|^2 x d\vec{x}_\perp dt' \nonumber \\
  &+& \frac{\varepsilon_0 c}{2k_0} \int {\rm Im}\left\{\psi^\star \partial_x \psi\right\}[t'-\verb"t"'_{\rm CE}(z)] d\vec{x}_\perp dt'.
\end{eqnarray}
The intrinsic transverse OAM is then evaluated as $J_y^{(i)} = J_y- J_y^{(e)}$, yielding
\begin{equation}\label{JYIO}
J_y^{(i)}=-\frac{\varepsilon_0 c}{2k_0}\int {\rm Im}\left\{\psi^\star \partial_x \psi\right\} [t'-t'_{\rm CE}(z)]d\vec{x}_\perp dt' .
\end{equation}
These are the expression provided in Ref. \cite{PORRAS_PIERS} for the transverse OAM and its parts. They can be further simplified if we restrict ourselves to propagation along $z$ with vanishing linear momentum along $x$:
\begin{equation}\label{PXO}
P_x=\int \langle T_{yz}\rangle d\vec{x}_\perp dt' =\frac{\varepsilon_0}{2k_0}\int {\rm Im}\left\{\psi^\star \partial_x \psi\right\}  d\vec{x}_\perp dt'=0 ,
\end{equation}
and similarly for $P_y$ (this excludes slightly inclined propagation allowed in the paraxial approximation). Then Eqs. (\ref{JYO},\ref{JYEO},\ref{JYIO}) simplify to
\begin{eqnarray}
J_y&=& -\frac{1}{2}\varepsilon_0 \int |\psi|^2 x d\vec{x}_\perp dt', \label{JYO2} \\
J_y^{(e)}&=&- \frac{1}{2}\varepsilon_0 \int |\psi|^2 x d\vec{x}_\perp dt' \!+\! \frac{\varepsilon_0 c}{2k_0}\!\int\! {\rm Im}\left\{\psi^\star \partial_x \psi\right\} t'd\vec{x}_\perp dt' ,\label{JYEO2} \\
J_y^{(i)}&=&-\frac{\varepsilon_0 c}{2k_0}\int {\rm Im}\left\{\psi^\star \partial_x \psi\right\} t'd\vec{x}_\perp dt' \label{JYIO2} .
\end{eqnarray}

The advantage of this formulation is that the determination of the transverse OAM requires the knowledge of the field only at a single transversal plane, as usually observed and retrieved from diverse temporally-resolved and spatially-resolved characterization techniques.

\section{Spatial formulation}

Designing or observing an optical wave packet in three dimensions $\psi(x,y,z',t)$ at an instant of time $t$ is not usual in experiments. This formulation may look intuitive since we are used to seeing mechanical objects evolving in time, but for light it is a rather theoretical  view.  Now, we conveniently evaluate the energy and transverse OAM as
\begin{equation}
W=\int \langle w\rangle d\vec x_\perp dz',\quad J_y = \int \langle j_y \rangle d\vec x_\perp dz',
\end{equation}
where we have taken into account that integration to all $z$ at given $t$ is the same as integration to all $z'$.

The energy density is $w=(\varepsilon_0 E_iE_i + \mu_0^{-1}B_iB_i)/2$ and the $y$ component of angular momentum density is $j_y=zp_x-xp_z$, where $p_x=S_x/c^2$ and $p_z=S_z/c^2$, and $S_x=\mu_0^{-1}(E_yB_z - E_zB_y)$ and $S_z=\mu_0^{-1}(E_xB_y-E_yB_x)$. Evaluation of their cycle-averaged values with either Eqs. (\ref{FIELDSX}) or (\ref{FIELDSY}) yield $\langle w\rangle=\varepsilon_0|\psi|^2/2$ and $W=\int \langle w \rangle dV$, or
\begin{equation}\label{EM}
W=\frac{\varepsilon_0}{2}\int |\psi|^2 d\vec x_\perp dz',
\end{equation}
and
\begin{equation}
  \langle j_y \rangle = z\langle p_x\rangle - x \langle p_z\rangle = (z'+ct) \langle p_x\rangle - x \langle p_z\rangle,
\end{equation}
where
\begin{equation}\label{PXPZ}
  \langle p_x\rangle = \frac{\varepsilon_0}{2\omega_0} {\rm Im}\{\psi^\star \partial_x\psi\}, \quad \langle p_z\rangle = \frac{\varepsilon_0}{2c}|\psi|^2,
\end{equation}
which determine the $y$ component of the about the origin O as
\begin{equation}\label{JYM}
J_y =\frac{\varepsilon_0}{2\omega_0} \int \left[(z'+ct){\rm Im} \{\psi^\star \partial_x\psi\} - k_0x|\psi|^2\right]d\vec x_\perp dz',
\end{equation}
which we identify with the transverse OAM.

Again, we can extract the extrinsic OAM of the STOV center with respect to the origin OAM as
\begin{equation}
J_y^{(e)} =  [ct+ z'_{\rm CE}(t)]\int \langle p_x\rangle d\vec x_\perp dz' - x_{\rm CE}(t) \int \langle p_z\rangle d\vec x_\perp dz',
\end{equation}
where
\begin{equation}\label{CENTERXP}
x_{\rm CE}(t) = \frac{1}{W} \int \langle w \rangle x d\vec x_\perp dz' = \frac{\int |\psi|^2 x d\vec x_\perp dz'}{\int|\psi|^2 d\vec x_\perp dz'},
\end{equation}
an analogous expression of $y_{\rm CE}(t)$, and
\begin{equation}\label{CENTERZP}
z'_{\rm CE}(t) = \frac{1}{W} \int \langle w \rangle z' d\vec x_\perp dz' = \frac{\int |\psi|^2 z' d\vec x_\perp dz'}{\int|\psi|^2 d\vec x_\perp dz'}
\end{equation}
determine the center of the energy density in three-dimensional space $(x,y,z')$, and should not be confused with $\verb"x"_{\rm CE}(z)$, $\verb"y"_{\rm CE}(z)$, and $c\verb"t"'_{\rm CE}(z)$. In particular $z'_{\rm CE}(t)$ accounts for the fact that the spatial center may not be at $z'=0$ for a wave packet lacking symmetries. Use of Eq. (\ref{PXPZ}) then yields
\begin{eqnarray}\label{JYEM}
  J_y^{(e)} &=& \frac{\varepsilon_0}{2\omega_0} \left\{ [ct + z'_{\rm CE}(t)] \!\int \! {\rm Im} \{\psi^\star \partial_x\psi\} d\vec x_\perp dz'\right. \nonumber \\
  & & \qquad\qquad \left.- k_0 x_{\rm CE}(t)\!\int \! |\psi|^2 d\vec x_\perp dz'\right\}.
\end{eqnarray}
The intrinsic OAM is then evaluated as $J_y^{(i)} =J_y - J_y^{(e)}$. After some simplification using Eq. (\ref{CENTERXP}), the result is

\begin{equation}\label{JYIM}
J_y^{(i)} =  \frac{\varepsilon_0}{2\omega_0} \int [z'-z'_{\rm CE}(t)] {\rm Im} \{\psi^\star \partial_x\psi\} d\vec x_\perp dz'.
\end{equation}
Again, these expressions can be further simplified if we restrict ourselves to propagation along $z$ with zero linear momentum along $x$ and $y$, a condition that now reads as
\begin{equation}\label{PXM}
P_x=\int \langle p_x \rangle d\vec x_\perp dz' =\frac{\varepsilon_0}{2\omega_0}\int {\rm Im}\left\{\psi^\star \partial_x \psi\right\}  d\vec x_\perp dz' =0 ,
\end{equation}
and analogously for $P_y$. Then the set of Eqs. (\ref{JYM},\ref{JYEM},\ref{JYIM}) simplify to
\begin{eqnarray}
J_y &=&\frac{\varepsilon_0}{2\omega_0} \int \left[z'{\rm Im} \{\psi^\star \partial_x\psi\} - k_0x|\psi|^2\right]d\vec x_\perp dz' ,\label{JYM2} \\
J_y^{(e)} &=& -\frac{\varepsilon_0}{2c} x_{\rm CE}(t)\int  |\psi|^2 d\vec x_\perp dz' ,\label{JYEM2}\\
J_y^{(i)} &=&  \frac{\varepsilon_0}{2\omega_0} \int z' {\rm Im} \{\psi^\star \partial_x\psi\} d\vec x_\perp dz' .\label{JYIM2}
\end{eqnarray}
This formulation is useful when an optical wave packet is specified in three dimensions at an instant of time, which may be of interest from a more theoretical perspective.

\section{Application to canonical STOVs}

Next we apply the two formulations above to the canonical STOVs considered in \cite{PORRAS_PIERS} and \cite{BLIOKH_PRA_2023,BLIOKH_PRL_2021}.

\subsection{Transverse OAM of elliptical STOVs in space-time}

\begin{figure}
\begin{center}
\includegraphics*[height=4cm]{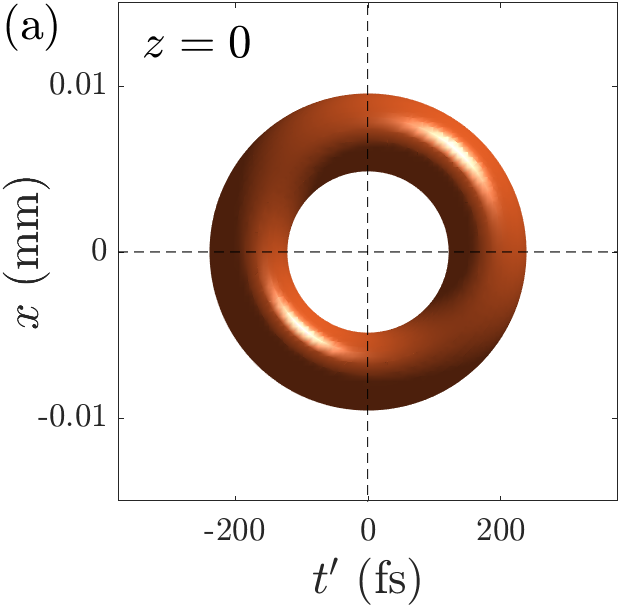}\hspace*{0.3cm}\includegraphics*[height=4cm]{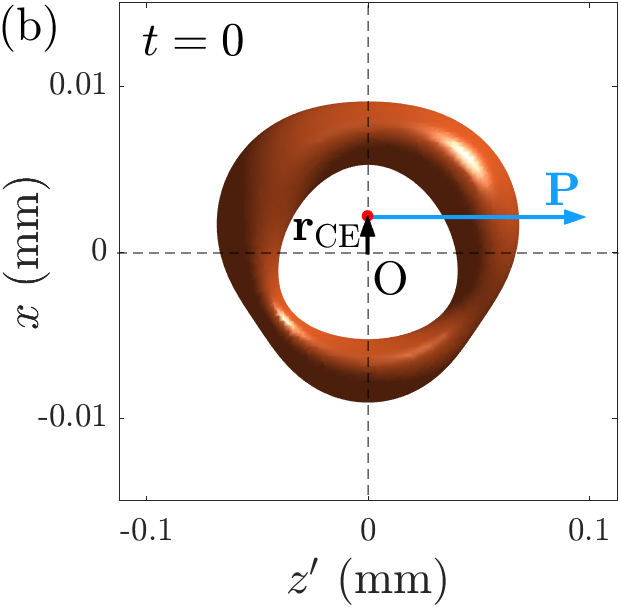}
\end{center}
\caption{\label{Fig2} Iso-energy-density surfaces ($80\%$ of the maximum value) (a) in $(x,y,t')$ at $z=0$ and (b) in $(x,y,z')$ at the time of arrival $t=0$ at $z=0$ of the STOV in Eq. (\ref{STOVZ}) with $\omega_0=2.8$ rad/fs, $l=1$, $w_0=0.01$ mm, and $t_0=250$ fs (ellipticity $\gamma=ct_0/w_0=7.5$). The $y$-axis is outward from the screen.}
\end{figure}

An elliptical STOV in space-time at a transversal plane is of the form $\psi= f(\varrho)e^{-il\varphi} g(y)$, where $\varrho=\sqrt{\tau^2+\xi^2}$, $\varphi={\rm tan}^{-1}(\xi/\tau)$ are spatiotemporal polar coordinates for the scaled time $\tau =t'/t_0$ and transversal coordinate $\xi=x/x_0$. The function $g(y)$ is an arbitrary profile along $y$. The parameters $t_0$ and $x_0$ determine the duration, the transversal size, and the ellipticity $\gamma=ct_0/x_0$. The simplest elliptical STOV in space-time is
\begin{equation}\label{STOV0}
  \psi(x,y,t') = e^{-\frac{y^2}{w_0^2}}e^{-\frac{x^2}{w_0^2}} e^{-\frac{t^{\prime 2}}{t_0^{2}}}\left(\frac{t^{\prime}}{t_0}- i\mbox{sign}(l) \frac{x}{w_0}\right)^{|l|}\!\!,
\end{equation}
for which $f(\varrho)=\varrho^{|l|}e^{-\varrho^2}e^{-il\varphi}$. The elliptical shape of the energy density in $(x,y,t')$ is illustrated in Fig. \ref{Fig2}(a) with a surface of constant energy density for a particular choice of the parameters. For these STOVs, Eq. (\ref{EO}) yields
\begin{equation}
W=\varepsilon_0c\pi t_0 x_0 \int_0^\infty d\varrho \varrho|f(\varrho)|^2 \int |g(y)|^2 dy .
\end{equation}
Equation (\ref{PXO}) leads to $P_x=0$, and Eqs. (\ref{JYO2},\ref{JYEO2},\ref{JYIO2}), conveniently rewritten in spatiotemporal polar coordinates, results in $J_y=0$, and
\begin{equation}
J_y^{(e)}= - J_y^{(i)}= l\frac{\varepsilon_0c}{2 k_0} t_0^2\pi \int_0^\infty d\varrho \varrho|f(\varrho)|^2 \int |g(y)|^2 dy,
\end{equation}
from which we recover the result in Ref. \cite{PORRAS_PIERS} that
\begin{equation}\label{RIGHT}
   \frac{J_y}{W} = 0 , \quad  \frac{J_y^{(e)}}{W}=-\frac{l}{2\omega_0}\gamma , \quad \frac{J_y^{(i)}}{W} =\frac{l}{2\omega_0}\gamma .
\end{equation}
STOVs that are elliptical in space-time at a transversal plane do not carry transverse OAM about the origin O because the extrinsic and the intrinsic transverse OAM are opposite.

Evaluation of the transverse OAM of the same elliptical STOV in space-time with the spatial formulation is not trivial since we need to know the full three-dimensional structure at least at an instant of time. Fortunately, this is possible half analytically and half numerically for the STOV in Eq. (\ref{STOV0}) using the propagation expressions provided in \cite{PORRAS_OL_2023}. Application of the spatial formulation will then teach us that this formulation provides the same results. In passing, it will reveal that the an elliptical STOV in space-time is not elliptical in three-dimensional space, which in turn sheds light on the origin of their extrinsic transverse OAM.

As detailed in the Appendix, the propagation of a paraxial ($|\partial_z\psi |\ll k_0|\psi|$) and quasi-monochromatic ($|\partial_{t'}\psi |\ll \omega_0|\psi|$) scalar field $\psi(x,y,t',z)e^{i(k_0z-\omega_0 t)}$ in free space can be described by the Schr\"odinger equation
\begin{equation}\label{SCH1}
\partial_z\psi =\frac{i}{2k_0}\Delta_\perp \psi .
\end{equation}
The solution of Eq. (\ref{SCH1}) with the STOV in Eq. (\ref{STOV0}) at $z=0$ was obtained from Fresnel diffraction integral in \cite{PORRAS_OL_2023} as
\begin{equation}\label{STOVZ}
\begin{split}
  &\psi(x,y,t',z) = \frac{-iz_R}{q(z)} e^{\frac{ik_0y^2}{2q(z)}} e^{\frac{ik_0x^2}{2q(z)}}e^{-\frac{t'^2}{t_0^2}} \left(\frac{z}{q(z)}\right)^{\frac{|l|}{2}} \\
  &\times   \frac{1}{2^{|l|}}H_{|l|}\!\left\{\!\left(\frac{q(z)}{z}\right)^{\frac{1}{2}}\!\!\left[\frac{t'}{t_0} - \mbox{sign}(l) \frac{x}{w_0}\frac{z_R}{q(z)} \right]\!\right\}
\end{split}
\end{equation}
where $q(z)=z-iz_R$, $z_R=k_0w_0^2/2$, and $H_{|l|}(\cdot)$ is the Hermite polynomial of order $|l|$. To proceed with the spatial formulation, we re-express Eq. (\ref{STOVZ}) replacing $z=z'+ct$ and $t'=-z'/c$ as
\begin{equation}\label{STOVZP}
\begin{split}
  &\psi(x,y,z',t) = \frac{-iz_R}{q(z',t)} e^{\frac{ik_0y^2}{2q(z',t)}} e^{\frac{ik_0x^2}{2q(z',t)}}e^{-\frac{z'^2}{c^2t_0^2}} \left(\frac{z'+ct}{q(z',t)}\right)^{\frac{|l|}{2}} \\
  &\times   \frac{1}{2^{|l|}}H_{|l|}\!\left\{\!\left(\frac{q(z',t)}{z'+ct}\right)^{\frac{1}{2}}\!\!\left[-\frac{z'}{ct_0} - \mbox{sign}(l) \frac{x}{w_0}\frac{z_R}{q(z',t)} \right]\!\right\} ,
\end{split}
\end{equation}
with $q(z',t)=z' + ct -iz_R$, which is exactly the same STOV as that in Eq. (\ref{STOVZ}) elliptical in $(x,y,t')$ at $z=0$. 

The lack of elliptical symmetry of the energy density in $(x,y,z')$ of the STOV in Eqs. (\ref{STOVZ}) or (\ref{STOVZP}) is evidenced in Fig. \ref{Fig2}(b), where the surface of the same constant value of the energy density of the same STOV as in Fig. \ref{Fig2}(a) at $z=0$ is shown at the instant of time $t=0$ of arrival at $z=0$. We have chosen a set of parameters that enhance the non-ellipticity, but as a matter of fact, an elliptical STOV in $(x,y,t')$ is never perfectly elliptical in $(x,y,z')$. This fact has not been noticed before, and evidences the existence of an extrinsic OAM. Clearly, the center of the energy in $(x,y,z')$ is not the origin O. For positive $l$, $x_{\rm CE}$ is positive, which yields a negative intrinsic transverse OAM. For negative $l$, the energy density is inverted  vertically and $x_{\rm CE}$ is negative, which yields a positive intrinsic transverse OAM. This all fits with Eqs. (\ref{RIGHT}) from the space-time formulation. Quantitatively, for the STOV in Fig. \ref{Fig2}, the center of the energy is numerically calculated to be $x_{\rm CE} = 0.4018\times 10^{-3}$ mm, $y_{\rm CE}=0$ mm, and $z'_{\rm CE}=0$ mm from Eqs. (\ref{CENTERXP},\ref{CENTERZP}) with Eq. (\ref{STOVZP}). The transverse OAM and its extrinsic and intrinsic parts are found to be, from Eqs. (\ref{EM}) and Eqs. (\ref{JYM2},\ref{JYEM2},\ref{JYIM2}) with Eq. (\ref{STOVZP}), $J_y/W= 0$ fs, $J_y^{(e)}/W=-1.339$ fs, and $J_y^{(i)}/W=1.339$ fs. These values coincide with those in Eqs. (\ref{RIGHT}) from the space-time formulation, since $(l/2\omega_0)\gamma= 1.339$ fs for this STOV.

In Ref. \cite{MILCHBERG_PRX} (Appendix A.1.), the authors attribute elliptical shape in space to elliptical STOVs in space-time. This misconception led the authors to the wrong conclusion that the extrinsic transverse OAM about the origin O vanishes.

\subsection{Transverse OAM of elliptical STOVs in space}

\begin{figure}
\begin{center}
\includegraphics*[height=4cm]{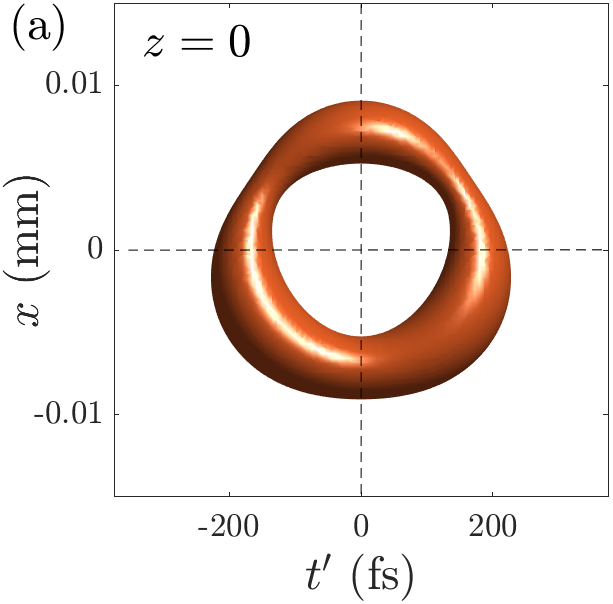}\hspace*{0.3cm}\includegraphics*[height=4cm]{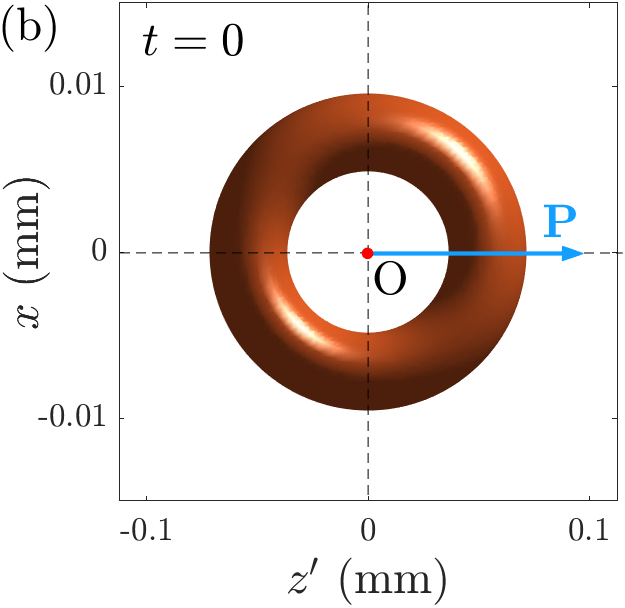}
\end{center}
\caption{\label{Fig3} Iso-energy-density surfaces ($80\%$ of the maximum value) (a) in $(x,y,z')$ at $t=0$ and (b) in $(x,y,t')$ at the plane $z=0$ where it is at $t=0$, of the STOV in Eq. (\ref{STOVZM}) with $\omega_0=2.8$ rad/fs, $l=1$, $w_0=0.01$ mm, and $z_0=0.075$ mm (ellipticity $\gamma=z_0/w_0=7.5$). The $y$-axis is outward from the screen.}
\end{figure}

An elliptical STOV in space at an instant of time is of the form $\psi=f(\rho) e^{il\phi}g(y)$, where $\rho=\sqrt{\zeta^2 +\xi^2}$ and $\phi={\rm tan}^{-1}(\xi/\zeta)$ are polar coordinates for the scaled axial coordinate $z'/z_0$ and transverse coordinate $x/x_0$. The parameters $z_0$ and $x_0$ determine the dimensions and the ellipticity $\gamma=z_0/x_0$. An example is
\begin{equation}\label{STOV0M}
\psi(x,y,z') = e^{-\frac{y^2}{w_0^2}}e^{-\frac{x^2}{w_0^2}} e^{-\frac{z^{\prime 2}}{z_0^{2}}}\left(\frac{z^{\prime}}{z_0} + i\mbox{sign}(l) \frac{x}{w_0}\right)^{|l|}\!\!,
\end{equation}
In Fig. \ref{Fig3}(b) we observe the elliptical shape of the energy density in space. By purpose the same wavelength, topological charge and  ellipticity as the STOV of Fig. \ref{Fig2} are chosen. For these STOVs, Eq. (\ref{EM}) yields
\begin{equation}
W= \varepsilon_0\pi z_0x_0 \int d\rho \rho |f(\rho)|^2 \int dy |g(y)|^2.
 \end{equation}
Equation (\ref{PXM}) leads to $P_x=0$, and Eqs. (\ref{JYM2},\ref{JYEM2},\ref{JYIM2}), after being rewritten in polar coordinates, to $J_y^{(e)}=0$ and
\begin{equation}
J_y= J_y^{(i)}= l\frac{\varepsilon_0}{2 \omega_0} z_0^2\pi \int_0^\infty d\rho \rho|f(\rho)|^2 \int |g(y)|^2 dy.
\end{equation}
The composition of the transverse OAM per unit energy of elliptical STOVs in space is then
\begin{equation}\label{RIGHT2}
   \frac{J_y}{W} =\frac{l}{2\omega_0}\gamma , \quad  \frac{J_y^{(e)}}{W}=0, \quad \frac{J_y^{(i)}}{W} =\frac{l}{2\omega_0}\gamma ,
\end{equation}
The intrinsic transverse OAM coincides with the transverse OAM about the origin O, because the extrinsic OAM vanishes, as is evident from Fig. \ref{Fig3}(b). This is the result obtained in Ref. \cite{BLIOKH_PRA_2023} for Bessel-type STOVs when the center of the STOV is chosen as the center of the energy. In Ref. \cite{MILCHBERG_PRX}, these values are attributed to elliptical STOVs in space-time.

Evaluation of the transverse OAM of the same elliptical STOV in space with the space-time formulation requires the knowledge of the structure of the STOV at all times. Even for the Bessel-type STOV in Ref. \cite{BLIOKH_PRL_2021}, the evolution in time is expressed as an integral that cannot be performed analytically. Here we resort on the dual vision between the evolution in $z$ of a wave packet specified as $\psi(x,y,t')$ at a transversal plane and the evolution in $t$ of a wave packet specified as $\psi(z,y,z')$ at a time, which allows us to find closed-form analytical expressions of the STOV in Eq. (\ref{STOV0M}) at all times. We then verify again that the space-time and spatial formulations provide the same results, and, on the way, that elliptical STOVs in space are no longer elliptical in space-time.

As shown in the Appendix, the evolution in time $t$ of a paraxial and quasi-monochromatic ( $|\partial_{z'}\psi |\ll k_0|\psi|$, $|\partial_t\psi |\ll \omega_0|\psi|$) scalar field $\psi(x,y,z',t)e^{i(k_0z-\omega_0 t)}$ in free space can be described by the Schr\"odinger equation
\begin{equation}\label{SCH2}
\partial_t\psi =\frac{ic}{2k_0}\Delta_\perp \psi ,
\end{equation}
which is dual to the Schr\"odinger equation (\ref{SCH1}) where $z$ is exchanged with $ct$. This symmetry allows us to write the temporal evolution of the STOV in Eq. (\ref{STOV0M}) at $t=0$ as governed by the Schr\"odinger equation (\ref{SCH2}) by interchanging $z\rightarrow ct$, $t'\rightarrow z'/c$, and $t_0\rightarrow z_0/c$ in the solution in Eq. (\ref{STOVZ}) of Schr\"odinger equation (\ref{SCH1}) with the STOV in Eq. (\ref{STOV0}) at $z=0$, which results in
\begin{equation}\label{STOVZM}
\begin{split}
  &\psi(x,y,z',t) = \frac{-it_R}{p(t)} e^{\frac{ik_0y^2}{2cp((t))}} e^{\frac{ik_0x^2}{2cp(t)}}e^{-\frac{z'^2}{z_0^2}} \left(\frac{t}{p(t)}\right)^{\frac{|l|}{2}} \\
  &\times   \frac{1}{2^{|l|}}H_{|l|}\!\left\{\!\left(\frac{p(t)}{t}\right)^{\frac{1}{2}}\!\!\left[\frac{z'}{z_0} + \mbox{sign}(l) \frac{x}{w_0}\frac{t_R}{p(t)} \right]\!\right\},
\end{split}
\end{equation}
where $p(t)=t-it_R$, and $t_R=k_0w_0^2/2c$ is a characteristic time of diffraction. To apply the space-time formulation we next use that $z'=-c t'$ and $t=t'+z/c$ to re-express Eq. (\ref{STOVZM}) as a function of $z$ and the local time $t'$ as
\begin{equation}\label{STOVZM2}
\begin{split}
  &\psi(x,y,t',z) = \frac{-it_R}{p(t',z)} e^{\frac{ik_0y^2}{2cp(t',z)}} e^{\frac{ik_0x^2}{2cp(t',z)}}e^{-\frac{c^2 t'^2}{z_0^2}} \left(\frac{t'+\frac{z}{c}}{p(t',z)}\right)^{\frac{|l|}{2}} \\
  &\times   \frac{1}{2^{|l|}}H_{|l|}\!\left\{\!\left(\frac{p(t',z)}{t'+\frac{z}{c}}\right)^{\frac{1}{2}}\!\!\left[-\frac{ct'}{z_0} + \mbox{sign}(l) \frac{x}{w_0}\frac{t_R}{p(t',z)} \right]\!\right\},
\end{split}
\end{equation}
where $p(t',z)=t'+z/c-i t_R$, which is exactly the same STOV as in Eq. (\ref{STOVZM}). Figure \ref{Fig3}(a) evidences that the elliptical STOV in space is not elliptical when observed in space-time. The parameters are chosen to enhance the non-ellipticity, but an elliptical STOV in $(x,y,z')$ is never perfectly elliptical in $(x,y,t')$.
Integrals in Eqs. (\ref{JYO2},\ref{JYEO2},\ref{JYIO2}) with the field in Eq. (\ref{STOVZM2}) cannot be performed analytically at any transversal plane. For the STOV in Fig. \ref{Fig3} the transverse OAM and its extrinsic and intrinsic parts at the plane $z=0$ at which the STOV arrives at the time $t=0$ are numerically calculated from Eqs. (\ref{EO}) and (\ref{JYO2},\ref{JYEO2},\ref{JYIO2}) with Eq. (\ref{STOVZM2}) to be $J_y/W= 1.339$ fs, $J_y^{(e)}/W=0$ fs, and $J_y^{(i)}/W=1.339$ fs, which are the same as those provided by Eqs. (\ref{RIGHT2}) from the spatial formulation, since $(l/2\omega_0)\gamma= 1.339$ fs for this STOV.

\section{Choice of the center of the photon wave function as STOV center}

In \cite{BLIOKH_PRA_2023}, the center of the STOV is considered to be the center of the photon wave function, or photon centroid, instead of the center of the energy. For elliptical STOVs in space, the photon centroid is said to be displaced from the energy center $x_{\rm CE}=0$ by $x_{\rm PC} = l/2\gamma k_0$ \cite{BLIOKH_PRA_2023}. Then Eq. (\ref{JYEM2}) must be replaced with $J_y^{(e)\prime}=-x_{\rm PC} P_z$, or, for elliptical STOVs in space, $J_y^{(e)\prime}= -(l/2\gamma k_0) (\varepsilon_0\pi z_0x_0/c) \int_0^\infty d\rho \rho|f(\rho)|^2 \int |g(\rho)|^2 dy$, and $J_y^{(e)\prime}/W =(-l/2\omega_0)(1/\gamma)$. We then conclude that
\begin{equation}\label{RIGHT3}
   \frac{J_y}{W} =\frac{l}{2\omega_0}\gamma , \,  \frac{J_y^{(e)\prime}}{W}=-\frac{l}{2\omega_0}\frac{1}{\gamma}, \, \frac{J_y^{(i)\prime}}{W} =\frac{l}{2\omega_0}\left(\gamma +\frac{1}{\gamma}\right) ,
\end{equation}
recovering the results in \cite{BLIOKH_PRA_2023,BLIOKH_PRL_2021,BLIOKH_PRA_2012} when the different type of STOVs and choice of centroid are both taken into account.

The advantage of choosing the center of the energy as the STOV center is that $J_y$, $J_y^{(e)}$ and $J_y^{(i)}$ are all conserved \cite{PORRAS_PIERS}. In contrast, with the photon centroid $J_y^{(e)\prime}$ and $J_y^{(i)\prime}$ are not generally conserved, as illustrated by the following example.

Consider the tilted, lobulated field in space at $t=0$
\begin{equation}\label{LOBULATED}
  \psi(x,y,z')= e^{-\frac{y^2}{w_0^2}}e^{-\frac{x^2}{w_0^2}} e^{-\frac{z^{\prime 2}}{z_0^{2}}}\frac{1}{2^{|l|}} H_{|l|}\left(\frac{z'}{z_0} +{\rm sign}(l) \frac{x}{w_0}\right),
\end{equation}
which is the dual field of the lobulated field in space-time in Eq. (10) of Ref. \cite{PORRAS_OL_2023} at $z=0$. Since the field in Eq. (\ref{LOBULATED}) is real, $P_x=0$. For the same reason and by the symmetry of the field in Eq. (\ref{LOBULATED}), the transverse OAM in Eq. (\ref{JYIM2}) yields $J_y=0$, and then $J_y^{(e)\prime}= -x_{\rm PC}(0) P_z$ and $J_y^{(i)\prime}= x_{\rm PC}(0) P_z$. We do not know $x_{\rm PC}(0)$ but given the exponential localization, it must take a finite value.

The propagated field of the dual lobulated field in space-time along $z$ as solution of the Schr\"odinger equation (\ref{SCH1}) was obtained also in Ref. \cite{PORRAS_OL_2023}. The replacements of $z\rightarrow ct$, $t'\rightarrow z'/c$, $t_0\rightarrow z_0/c$ in Eq. (13) of Ref. \cite{PORRAS_OL_2023} for that propagated field allows us to write the solution of the Schr\"odinger equation (\ref{SCH2}) with Eq. (\ref{LOBULATED}) at the initial time $t=0$ as
\begin{equation}\label{LOBULATEDZ}
\begin{split}
  &\psi(x,y,z',t)= \frac{-it_R}{p(t)} e^{\frac{ik_0y^2}{2cp(t)}}e^{\frac{ik_0x^2}{2cp(t)}} e^{-\frac{z^{\prime 2}}{z_0^2}}\left(\frac{-it_R}{p(t)}\right)^{\frac{|l|}{2}}\\
  &\times \frac{1}{2^{|l|}} H_{|l|} \left\{\left(\frac{p(t)}{-it_R}\right)^{\frac{1}{2}} \left[\frac{z'}{z_0} + {\rm sign}(l)\frac{x}{w_0}\left(\frac{-it_R}{p(t)}\right)\right] \right\},
\end{split}
\end{equation}
where $p(t)=t-it_R$ and $t_R=k_0w_0^2/2c$ as before. At long times $t \pm \infty$, the lobulated field tends to form the elliptical STOVs in space
\begin{eqnarray}
\psi(x,y,z',t)&\rightarrow& \frac{t_R}{|t|} e^{\frac{ik_0y^2}{ct}}e^{-\frac{y^2}{w^2(t)}}e^{\frac{ik_0x^2}{ct}} e^{-\frac{x^2}{w^2(t)}} e^{-\frac{z^{\prime 2}}{z_0^2}}\nonumber \\
&\times& \frac{1}{2^{|l|}}\!\left\{\begin{array}{ll} \!(-i)\left(\frac{z'}{z_0}\!-\!i{\rm sign}(l) \frac{x}{w(t)}\right)^{|l|}  & t\uparrow +\infty\\
                                               (+i)\!\left(\frac{z'}{z_0}\!+\!i{\rm sign}(l) \frac{x}{w(t)}\right)^{|l|}  & t\downarrow -\infty \end{array} \right.
\end{eqnarray}
where $w(t)=w_0 |t|/t_R$ growing linearly with time, and of ellipticity $\gamma = z_0/w(t)$ going to zero. Since the field approaches elliptical STOVs of opposite topological charge $-l$ for $t\rightarrow\infty$ and $+l$ for $t\rightarrow -\infty$, $x_{\rm PC}(t)$ must approach $l/2\gamma k_0$, which in turn tends to $-\infty$ for $t\rightarrow +\infty$, and to $+\infty$ for $t\rightarrow -\infty$. Consequently, $J_y^{(e)\prime}=-x_{\rm CE}(t)P_z$ tends to $+\infty$ for $t\rightarrow +\infty$ and to $-\infty$ for $t\rightarrow -\infty$, since $P_z$ is constant and positive, and vice versa for $J_y^{(i)\prime}=x_{\rm CE}(t)P_z$, being therefore non-conserved and approaching opposite infinitely large values. In contrast, Eqs. (\ref{JYM2},\ref{JYEM2},\ref{JYIM2}) with the center of the energy $x_{\rm CE}=0$ yield $J_y=J_y^{(e)}=J_y^{(i)}=0$.

\section{Conclusions}

In short, we have shown that the controverted results for the transverse OAM of STOVs in Refs. \cite{BLIOKH_PRA_2023,BLIOKH_PRL_2021, BLIOKH_PRA_2012} and \cite{PORRAS_PIERS} are all correct as they refer to different STOVs and different choices of the STOV center. The theory presented here in its two equivalent formulations accounts for all these correct values. The theory presented in \cite{HANCOCK_PRL} and the arguments in \cite{MILCHBERG_PRX} against \cite{PORRAS_PIERS} and \cite{BLIOKH_PRA_2023} are wrong since the space-time view and the spatial view are mixed when assuming elliptical STOVs in space-time are also elliptical in space.

The space-time formulation, and in particular Eqs. (\ref{JYO2},\ref{JYEO2}\ref{JYIO2}), being closer to the experiments, and also to most of the theories on ultrashort, spatiotemporal light fields, provide the natural frame for a correct evaluation of the transverse OAM, and the reference values of the canonical STOVs that are purportedly generated in most of the experiments.

\section*{Funding}
Ministerio de Ciencia e Innovación (PID2021-122711NB-C21).

\section*{Acknowledgments}
This work has been partially supported by the Spanish Ministry of Science and Innovation, Gobierno de España, under Contract No. PID2021-122711NB-C21.

\appendix

\section{Schr\"odinger equations for axial and temporal propagation of paraxial and quasi-monochromatic scalar fields}

We consider the wave equation
\begin{equation}\label{WAVE}
  \Delta E -\frac{1}{c^2}\frac{\partial^2 E}{\partial t^2} =0
\end{equation}
for a scalar field $E$ in free space.

\subsection{Space-time formulation}

In this formulation we change to the variable $t'=t-z/c$ and $z$ is unchanged, with which the wave equation reads as
\begin{equation}
  \Delta_\perp E + \frac{\partial^2 E}{\partial z^2} - \frac{2}{c} \frac{\partial^2 E}{\partial z \partial t'}=0 .
\end{equation}
For a field of the form $E=\psi e^{-i\omega_0 t'}$, the above equation becomes
\begin{equation}\label{PASO1}
\Delta_\perp \psi + \frac{\partial}{\partial z}\left[\frac{\partial \psi}{\partial z} - \frac{2}{c}\left(\frac{\partial\psi}{\partial t'}- i\omega_0 \psi\right)\right]=0.
\end{equation}
For a paraxial and quasi-monochromatic field, the changes in the envelope on propagation are much slower than the axial oscillations at $k_0$, and in time much slower than the temporal oscillations at $\omega_0$, implying that $|\partial\psi/\partial z|\ll k_0|\psi|$ and $|\partial\psi/\partial t'|\ll \omega_0|\psi|$. Then Eq. (\ref{PASO1}) simplifies to
\begin{equation}\label{SCH3}
\Delta_\perp \psi + 2i k_0 \frac{\partial \psi}{\partial z}=0 ,
\end{equation}
which is Eq. (\ref{SCH1}). The solutions in the main text were obtained in Ref. \cite{PORRAS_OL_2023} from Fresnel diffraction integral
\begin{equation}\label{FRESNEL1}
  \psi(x,y,t',z)=\frac{k_0}{2\pi i z} \int d \vec x_\perp^{\prime} \psi(x,y,t') e^{\frac{ik_0}{2z}[(x-x')^2+(y-y')^2]}
\end{equation}
for a field specified in $(x,y,t')$ at $z=0$.

\subsection{Spatial formulation}

Now we change to the variables $z'=z-ct$ and $t$ remains unchanged, with which the wave equation reads as
\begin{equation}
  \Delta_\perp E -\frac{1}{c^2} \frac{\partial^2 E}{\partial t} + \frac{2}{c} \frac{\partial^2 E}{\partial z' \partial t}=0 .
\end{equation}
For a field of the form $E=\psi e^{k_0 z'}$, we obtain
\begin{equation}\label{PASO2}
\Delta_\perp \psi -\frac{1}{c^2} \frac{\partial}{\partial t}\left[\frac{\partial \psi}{\partial t} - 2c\left(\frac{\partial\psi}{\partial z'}+ ik_0 \psi\right)\right]=0.
\end{equation}
For a paraxial and quasi-monochromatic field, the changes in the envelope in its temporal evolution are much slower than the temporal oscillations at $\omega_0$ and in $z'$ much slower than the axial oscillations at $k_0$, implying that $|\partial\psi/\partial z'|\ll k_0|\psi|$ and $|\partial\psi/\partial t|\ll \omega_0|\psi|$. Then Eq. (\ref{PASO2}) simplifies to
\begin{equation}\label{SCH4}
\Delta_\perp \psi + 2i \frac{k_0}{c} \frac{\partial \psi}{\partial t}=0 ,
\end{equation}
which is Eq. (\ref{SCH2}) and the same as Eq. (\ref{SCH1}) replacing $z\rightarrow ct$. The solutions in the main text for a field specified in space at $(x,y,z')$ at $t=0$ can be obtained by also replacing $z\rightarrow ct$ in Fresnel diffraction integral (\ref{FRESNEL1}), i.e.,
\begin{equation}\label{FRESNEL2}
  \psi(x,y,z',t)=\frac{k_0}{2\pi i c t} \int d \vec x'_\perp \psi(x,y,z') e^{\frac{ik_0}{2ct}[(x-x')^2+(y-y')^2]}.
\end{equation}
In practice, the solutions $\psi(x,y,z',t)$ of (\ref{FRESNEL2}) in the main text are directly obtained by replacing $z\rightarrow ct$ in the solutions $\psi(x,y,t',z)$ of Eq. (\ref{FRESNEL1}).

Any solution $\psi(x,y,z',t)$ of Schr\"odinger equation (\ref{SCH4}) is also a solution of Schr\"rodinger equation (\ref{SCH3}) and vice versa: Changing to $t'=-z'/c$ and $z=z'+ct$ in Eq. (\ref{SCH4}), it follows that $\partial \psi/\partial t=c\partial \psi/\partial z$, and therefore Eq. (\ref{SCH3}). Thus the two Schr\"odinger equations (\ref{SCH3}) and (\ref{SCH4}) describe exactly the same phenomena, diffraction under paraxial and quasi-monochromatic conditions. They just pose and solve different problems: the evolution in $z$ of a wave packet specified as $\psi(x,y,t',z)$ at given $z$, and the evolution in $t$ of a wave packet specified as $\psi(x,y,z',t)$ at a given $t$. All scalar fields in Eqs. (\ref{STOVZ}, \ref{STOVZP}, \ref{STOVZM}, \ref{STOVZM2}, \ref{LOBULATEDZ}) in the main text satisfy both Schr\"odinger equations.

%%%%%%%%%%%%%%%%%%%%%%% References %%%%%%%%%%%%%%%%%%%%%%%%%

\end{document}